# InMRAM: Introductory course on Magnetic Random Access Memories for microelectronics students and engineers

Gregory Di Pendina, Guillaume Prenat, Bernard Dieny
Univ. Grenoble Alpes, INAC-SPINTEC, F-38000 Grenoble, France
CNRS, INAC-SPINTEC, F-38000Grenoble, France
CEA, INAC-SPINTEC, F-38000 Grenoble
bernard.dieny@cea.fr

*Abstract*— **Magnetic Random Access Memories (MRAM) interest is growing fast in the microelectronics industry. Commercial MRAM products already exist and all major industrial players have launched large R&D efforts to bring Spin Transfer Torque MRAM to production. The principle goal is to replace part of the memory hierarchy (in particular DRAM) by MRAM below the 20nm node. Interest in bringing non-volatile elements closer to the logic is also growing, in order to improve the performance of electronic circuits by increasing the bandwidth between logic and memory and reducing the power consumption. Despite this rising interest, very few microelectronics engineers have a background in magnetism and even more in spinelectronics so that it is very difficult for them to get into this emerging field. Magnetism and microelectronics communities have worked separately so far. We are convinced it is time to develop more relationships between these two communities through joint symposia, workshops, conferences, as well as to introduce some courses in magnetism and spinelectronics in the education of young microelectronics engineers. In this context, SPINTEC, a French lab associated to CEA/CNRS/Grenoble University, started organizing last year an annual Introductory Course on MRAM (InMRAM) for students, engineers and researchers having no or little background in magnetism. The second edition will take place in Grenoble from 2nd to 4th July 2014.**

## I. INTRODUCTION

Spinelectronics is a merging between magnetism and electronics. It started in 1988 with the discovery of Giant Magnetoresistance (GMR) [1]. The first application of GMR has been as very sensitive magnetoresistive sensors called spin-valves [2] which have been used as read heads in Hard Disk Drives between 1996 and 2004. Two other major breakthroughs have further stimulated this field. The first one was the observation of large tunnel magnetoresistance (TMR) at room temperature in magnetic tunnel junctions (MTJ), first in amorphous Alumina based MTJ (TMR amplitude in the range 40-70%) [3, 4] and more recently in crystalline MgO based MTJ (TMR amplitude in the range 100-600%) [5, 6]. MTJs consist of two magnetic metallic layers separated by a thin tunneling barrier. When a bias voltage is applied across the stack, the electron tunneling probability depends on the relative orientation of the magnetization in the two adjacent magnetic layers. In most common cases, the tunneling probability is higher when the magnetizations are parallel (yielding a minimum resistance) and lower when they are antiparallel (yielding a higher resistance). Somehow, the magnetic layers play the role of polarizer and analyzer for the spin of the electrons. MTJs offer very nice possibilities of integration with CMOS components for several reasons. Their resistance can be tuned by adjusting the thickness of the tunnel barrier to be of the order of a few kΩ i.e. compatible with the typical resistance of MOSFET. Furthermore, they can be grown above the CMOS technological levels and electrically connected to the CMOS components through Cu via. In this integration, there is no contamination of the CMOS components. The MTJs only require annealing at temperature of the order of 300°C compatible with the CMOS process. The MTJ can be used as variable resistance controlled by magnetic field or by current/voltage thanks to the spin transfer torque phenomenon (see further).

Soon after this discovery of large TMR at room temperature, it was realized that MTJ could be used to make arrays of memory elements called MRAM (Magnetic Random Access Memories). A significant R&D effort then started on MRAM written by pulses of magnetic field, mostly led by Freescale/Everspin. This company successfully launched a 4Mbit MRAM product in 2006 which is used in microcontrollers, automotive applications or space applications (hardness to radiations). However, this technology is not scalable beyond the 65nm technology node because of the large current required to generate the pulses of magnetic field during write.

This issue has been circumvented thanks to a second major discovery: the spin transfer torque (STT) phenomenon [7, 8]. This is the action that a spin polarized current exerts on the magnetization of a magnetic nanostructure due to the exchange interaction between the spin of the conduction electrons and the spin of the electrons responsible for the local magnetization. If the current density is large enough, this phenomenon can be used to switch the magnetization of a magnetic nanostructure. It is particularly useful in the context of MRAM to switch the magnetization of the storage layer. This phenomenon has been a major discovery because it provides a new way to manipulate a magnetization with much less energy and in a much more localized way than with a magnetic field [9]. STT induced magnetization switching has been first observed in fully metallic magnetic pillars [9] but also, after some optimization, in MTJs [10]. In STT, the magnetization switching is determined by the current density flowing through the MTJ. As a result, the current required to write in STT MRAM scales as the MTJ area down to very small values of the order of 10µA where it reaches a bottom limit set by the memory retention. This is very favorable for the downsize scalability of MRAM based on STT switching. This property renewed the interest for MRAM since the possibility for this technology to be scalable to and beyond the 20nm node then became clearer.

Now most of the major companies in microelectronics have launched significant R&D efforts in STT MRAM (Samsung, Hynix/Toshiba, IBM, TDK, NEC, Qualcomm, TSMC…) and a number of start-up companies have emerged in this field (Grandis subsequently acquired by Samsung, Crocus Technology, Spin Transfer Technology, Avalanche…). The interest for MRAM has been strongly rising over the past few years as illustrated in Figure 1 which represents the number of papers and patents published per year containing the word "MRAM". These data are derived from the scirus database (www.scirus.com).

MRAM have now a clear place in the ITRS. The ITRS-ERD and ERM working groups have also published a report in July 2010 in which "STT MRAM and Redox RAM were identified as emerging memory technologies recommended for accelerated R&D leading to scaling and commercialization of non-volatile RAM to and beyond the 16nm generation" [11]. The goal is here to replace parts of the memory hierarchy by non-volatile MRAM as illustrated in Fig.2. This would have lot of advantages in terms of reducing the power consumption (both static and dynamic), reducing the footprint on the wafer, speeding up the communication between logic and memory functions, introducing ultrafast reprogrammability capabilities in electronic circuits. In a further step, the conventional memory hierarchy could even be revisited with new logic-in-memory architecture wherein the memory would be distributed with a much finer granularity within the logic circuits as illustrated in Figure 2. These architectures would allow a Normally-Off/Instant-On electronics with much reduced power consumption and faster bandwidth between logic and memory [12].

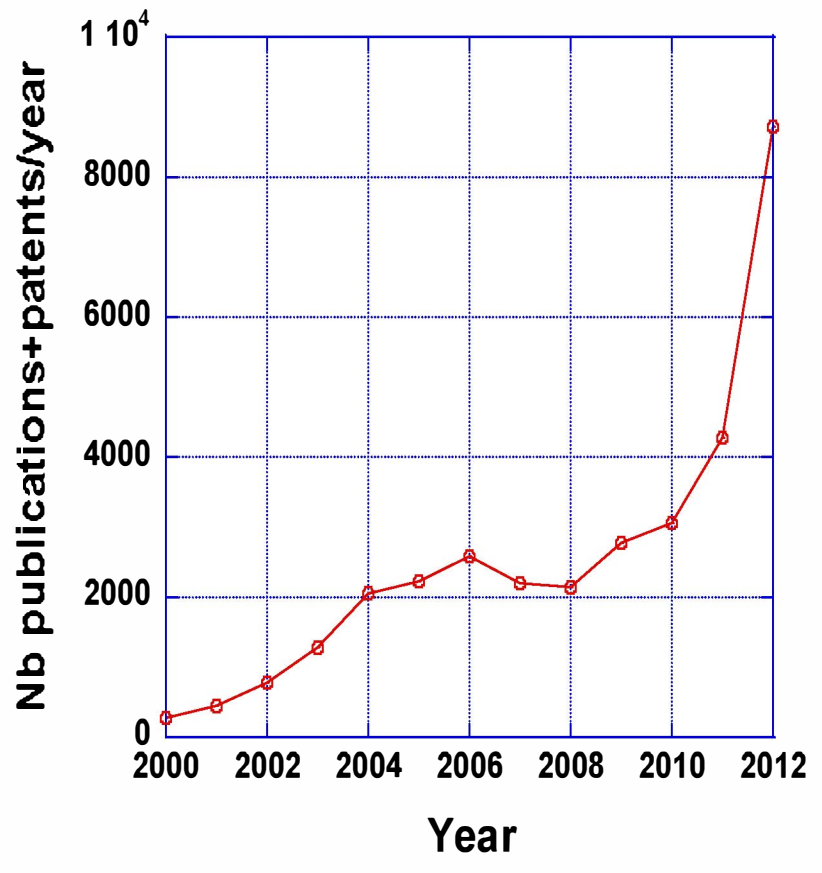

Fig. 1. Number of publications and patents per year containing the word "MRAM

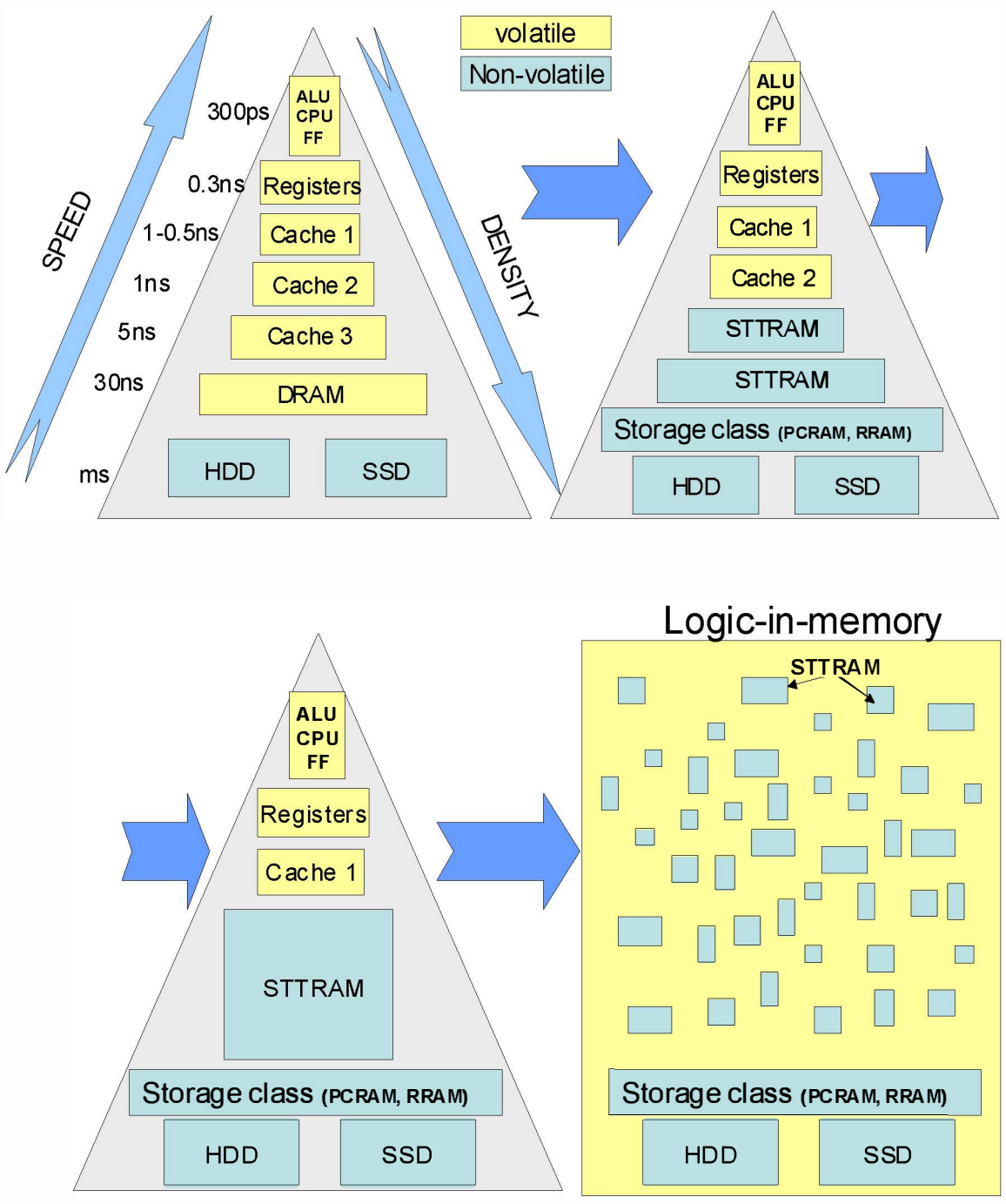

Fig. 2. Expected evolution of the memory hierarchy as the performances of MRAM will progress. As a first step, MRAM may replace DRAM below the 20nm node with cycle time of the order of 30ns. In a second step, as the switching speed of MRAM will increase for instance thanks to precessional STT switching, MTJ may be used even closer to the logic units. In a third step, new logic-in-memory architectures may be introduced wherein the memory is distributed with a much finer granularity within logic circuits.

Although most of the interest in MRAM is now focused on STT-MRAM, other companies in particular Crocus Technology are still developing MRAM based on field induced magnetization switching but in combination with thermal assistance. The idea is to take advantage of the Joule heating produced in the MTJ around the tunnel barrier when a current is flowing through it. At current densities of the order of a few $10^6$A/cm², the MTJ temperature can increase by 200°C in a few ns. By properly designing the MTJ stack, this temperature rise can be used to enable the magnetization switching with a weak magnetic field [13]. Thanks to this combination of field and heating, the field pulses can be shared between various bits (for instance writing all “0” in a 64 bit word with a single pulse of magnetic field, then all “1” with a second pulse), thereby strongly reducing the power consumption associated with field writing. This approach has very nice perspectives of applications in the field of security of data, but also for Content Addressable Memories (Fig.3) which are used in search engines, routers… [14]

Despite this rising interest for MRAM within the microelectronics industry, the field has mostly progressed until recently within the magnetism community. It turns out that for historical reasons, the R&D in magnetism and in microelectronics has developed separately. Magnetic materials have long been considered in microelectronic industry as impurities and carefully kept out of the fabs. Besides, the magnetism community has its own magnetism conferences (mostly MMM, Intermag, ICM) which gather most of the worldwide magnetism community. The microelectronics community has also a relatively large number of its own conferences, each of them addressing specific topics (basic issues, design, technology…). However, the two communities of magnetism and microelectronics have very few opportunities to meet and the cultural gap between these two R&D areas remains wide. In addition, since so far microelectronics engineers have received no or very little education in magnetism during their initial education, it is very difficult for them to understand talks and/or papers related to MRAM. The magnetic materials used in MRAM stacks look scaring with up to 15 stacked layers whose thickness must be controlled with an accuracy of 0.1nm for some of them. They seem almost impossible to manufacture from a common microelectronics point of view. However, these material developments have strongly benefited from the industrial know-how acquired in the past 15years thanks to the magnetic recording industry especially the development of spin-valves [2] and subsequently TMR [3-6] based read heads. Furthermore the spintronic phenomena used in MRAM and particularly STT-MRAM are quite complex to understand. Spin-dependent electron transport and magnetization dynamics phenomena are inter-dependent. This cultural gap and educational issues are actually hindering the development of MRAM technology. It is therefore important to start taking actions to foster more relationships between the two communities and introduce some courses in magnetism and spinelectronics within the initial education of microelectronic engineers.

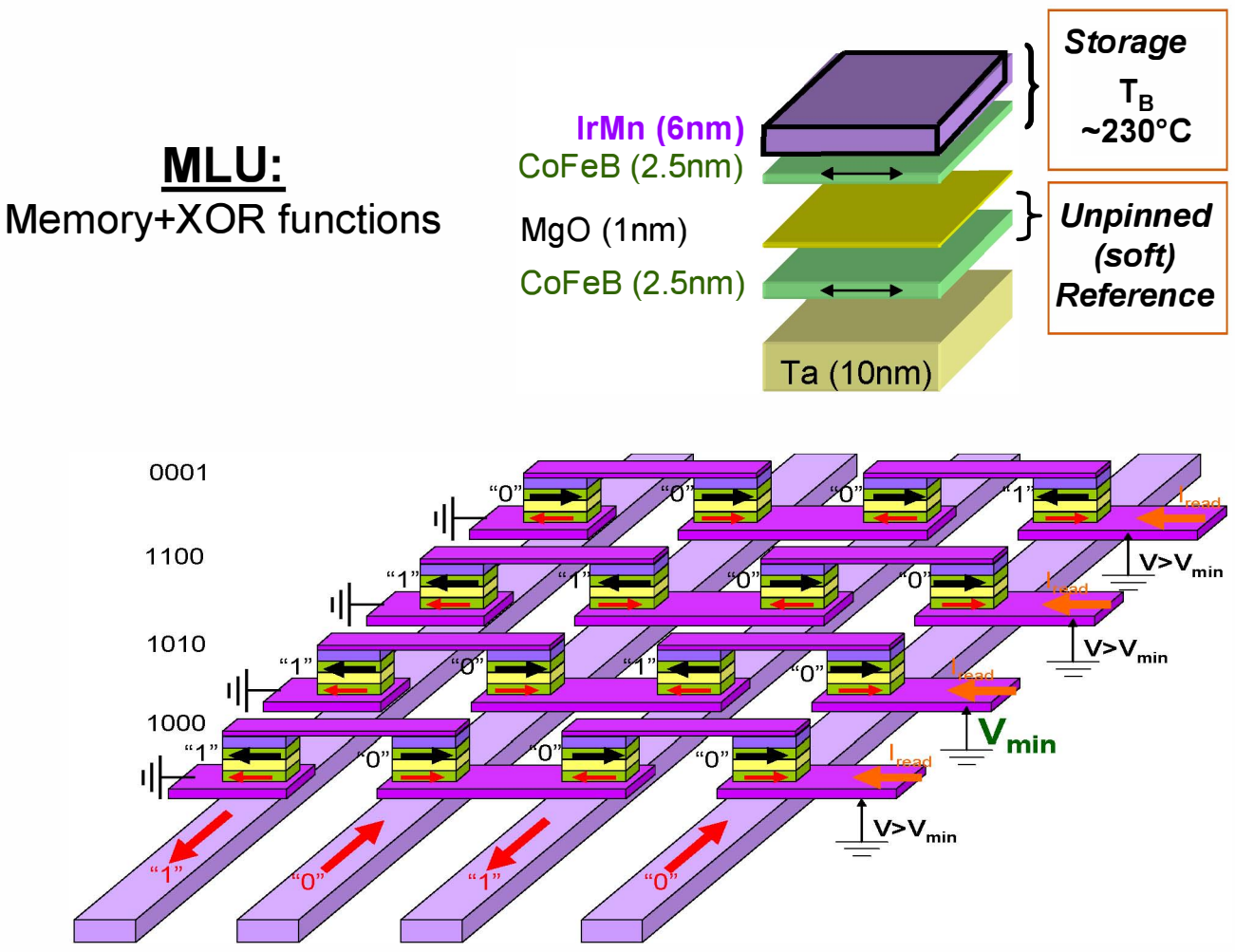

Fig. 3. concept of Magnetic Logic Unit (MLU) as developed by Crocus Technology (http://www.crocustechnology.com/pdf/Crocus_MIP_White Paper_v6.pdf) . A MLU is a MTJ comprising a soft reference layer in contrast to conventional MTJs which have a pinned reference layer. MLUs intrinsically combine a storage function together with a comparison (XOR) function. This is called the Match-In-Place function. MLUs are very useful for a number of applications in particular in NAND type of chains used for Content Addressable Memory (CAM) applications. The purpose of a CAM is to find at which address a particular word is located. The example above illustrates how to find the written word “1010” in a set of 4 addresses. By sending pulses of current corresponding to “1010” in the field generating lines located below the MTJ, the magnetization in all soft layers along each line get oriented along the corresponding field direction. Only the chain containing the searched word has the minimum resistance since all the MLU elements along this chain are in parallel magnetic configuration i.e. are in their minimum resistance state (3rd chain from the top in this example).

## II. How to favor links between magnetism and microelectronics communities ?

It is interesting to note that the places where the R&D in spinelectronics and MRAM has grown the fastest and the strongest are in places where expertise in both magnetism and microelectronics were present in close proximity. This is true for instance at Tohoku University in Japan or in Grenoble France. This is also true in companies which are producing or have produced both electronic devices and magnetic hard disk drives (Toshiba, Samsung, IBM, Hitachi…).

At the international level, it is important to organize joint events gathering magnetism and microelectronics experts. Several initiatives have been taken along this line in the past couple of years. For instance, a one-day workshop was organized in May 2010 at Hsinchu Taiwan, the eve of the start date of the VLSI and Intermag 2010 conferences which happened to take place the same week in Taiwan. This has been an excellent opportunity for comparing FLASH, ReRAM and MRAM technologies and getting reciprocally educated. Conferences such as IEDM, IMW, VLSI have clearly expressed their willingness to include more MRAM contributions in their technical program. So far, most of the MRAM reports are presented in magnetism conferences (Intermag and MMM). We should encourage the teams working on MRAM to also present their reports in microelectronics conferences. The IEEE Magnetic Society is very supportive of this endeavor. It has for instance sponsored a half-day short-course on MRAM at IMW2012. Reciprocally, it is important for magnetism experts working in MRAM and hybrid CMOS/Magnetic technology, who are most of the time physicists, to learn more about microelectronics so as to better orient their R&D efforts.

## III. InMRAM: In troductory Course on MRAM for students and engineers in Microelectronics

Magnetism is playing growing role in industry in general. Traditionally, the most important areas of use of magnetism has been in hard disk drives industry, in permanent magnets used for instance in electric motors, soft magnets used in inductors… in magnetic sensors implemented in a large variety of applications (automotive industry, robotics, electronic compass, …). Other areas of applications have more recently emerged related to biotechnology (magnetic particles for diagnostics or treatments, medical imaging…), or energy harvesting. Besides, MRAM and more generally hybrid electronic circuits combining CMOS and magnetic technologies is becoming a major direction of R&D at the merge between magnetism and microelectronics. Its potential of industrial development is huge.

In this regard, the worldwide number of engineers educated in magnetism and spinelectronics looks quite small. It seems time to introduce courses in magnetism and spinelectronics in electrical engineering education. This would give to the young engineers in this field the basis for understanding the materials, the phenomena, the various technical issues, related to MRAM technologies. To contribute in this direction, SPINTEC, which is a French lab affiliated to CEA (Commissariat à l'Energie Atomique et aux Energies Alternatives), CNRS (Centre National de la Recherche Scientifique) and Grenoble-University, organizes an annual Introductory Course on MRAM (InMRAM) whose first edition took place in 2013. The second edition will take place from July 2$^{nd}$ to 4$^{th}$ 2014 in Grenoble, France. This summer school addresses several topics related to magnetism, spinelectronics and MRAM technologies, as listed below:

- Magnetism basics for MRAM
  - ferromagnetism
  - antiferromagnetism
  - Interlayer coupling
  - Micromagnetism
  - walls and vortices
  - Superparamagnetic limit
  - exchange bias (ferro/antiferro coupling)
  - Anisotropy, in-plane/shape/perpendicular-to-plane anisotropy
  - Landau Lifshitz Gilbert magnetization dynamics equation
  - Magnetization switching by field
- Basic spinelectronic transport phenomena
  - Giant magnetoresistance (GMR)
  - Tunnel magnetoresistance (TMR)
  - Spin transfer torque (STT)
- Basics of MRAM
  - Growth and annealing of in-plane and out-of-plane magnetized MTJ
  - field induced switching MRAM (Stoner Wolfarth and toggle switching)
  - STT MRAM
  - Perpendicular-MTJ based STT MRAM
  - Thermally assisted MRAM
  - 3-terminal MRAM (Domain Wall and Spin-Orbit-Torque) MRAM
  - Comparison with other emerging non-volatile memory technologies
  - Benefits of MRAM compared to alternative technologies
  - Present and foreseen applications of MRAM: Microcontrollers, automotive applications, spatial applications, Embedded memories, DRAM replacement, SRAM replacement
- Magnetic back-end technology
  - Specific tools for back-end processing
  - MTJ deposition tools
  - Magneto-transport characterization
  - MTJ processing for MRAM
- Design tools for hybrid CMOS/MTJ technology
  - SPICE models for MTJ
  - layout and verification tools
  - Examples of circuit design
- Beyond MRAM
  - Combining the best of CMOS and magnetic worlds.
  - innovative architectures for non-volatile, reprogrammable logic, e.g., magnetic full adder, content addressable memory, M-FPGA, …
  - Logic-in-memory architecture
  - Normally-OFF/Instant-ON computing.
  - Low power electronics based on spintronics
  - Benefits from radiation hardness

InMRAM speakers are experts coming from Institut Néel, CNRS/Thales, CNRS, CEA, Yole and University of Montpellier 2. More details are given at www.inMRAM.com

The InMRAM first edition feedbacks from the 112 attendees were extremely positive. The attendees came from universities, research laboratories and companies as illustrated in Figure 4. This figure also shows that about 2/3 of the attendees were originating from the microelectronic community with no magnetism background and 1/3 from the magnetism community.

Concerning the geographical origin of the attendees, the latter was quite diverse. Approximately half of the audience was from France and the other half from 20 different countries, as illustrated in Figure 5.

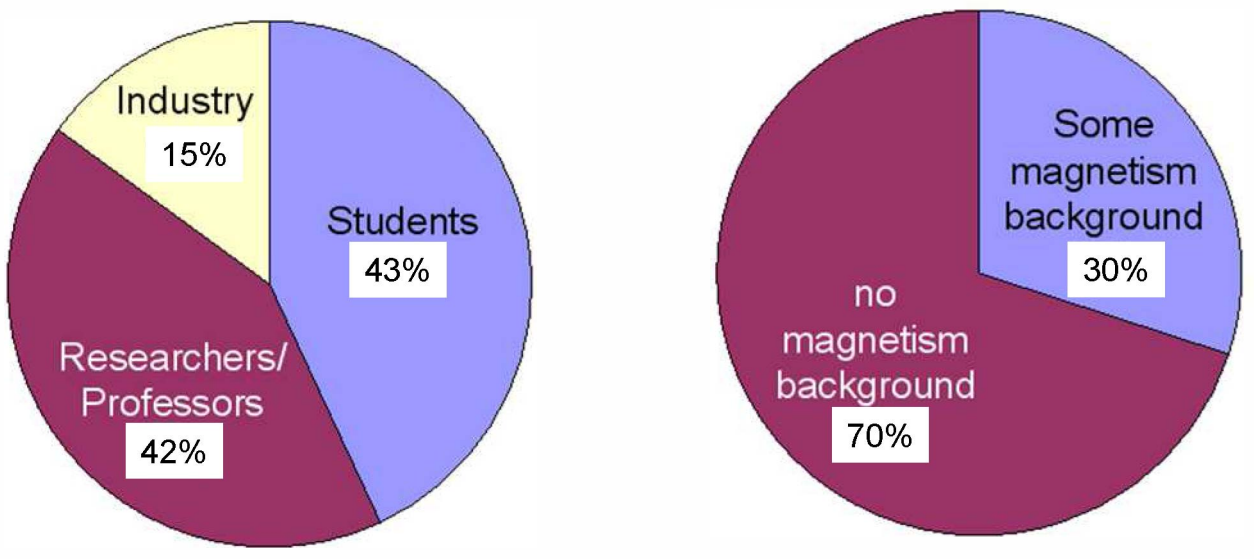

Fig. 4. Repartition of attendees in terms of (a) professional origin and (b) scientific/technical background.

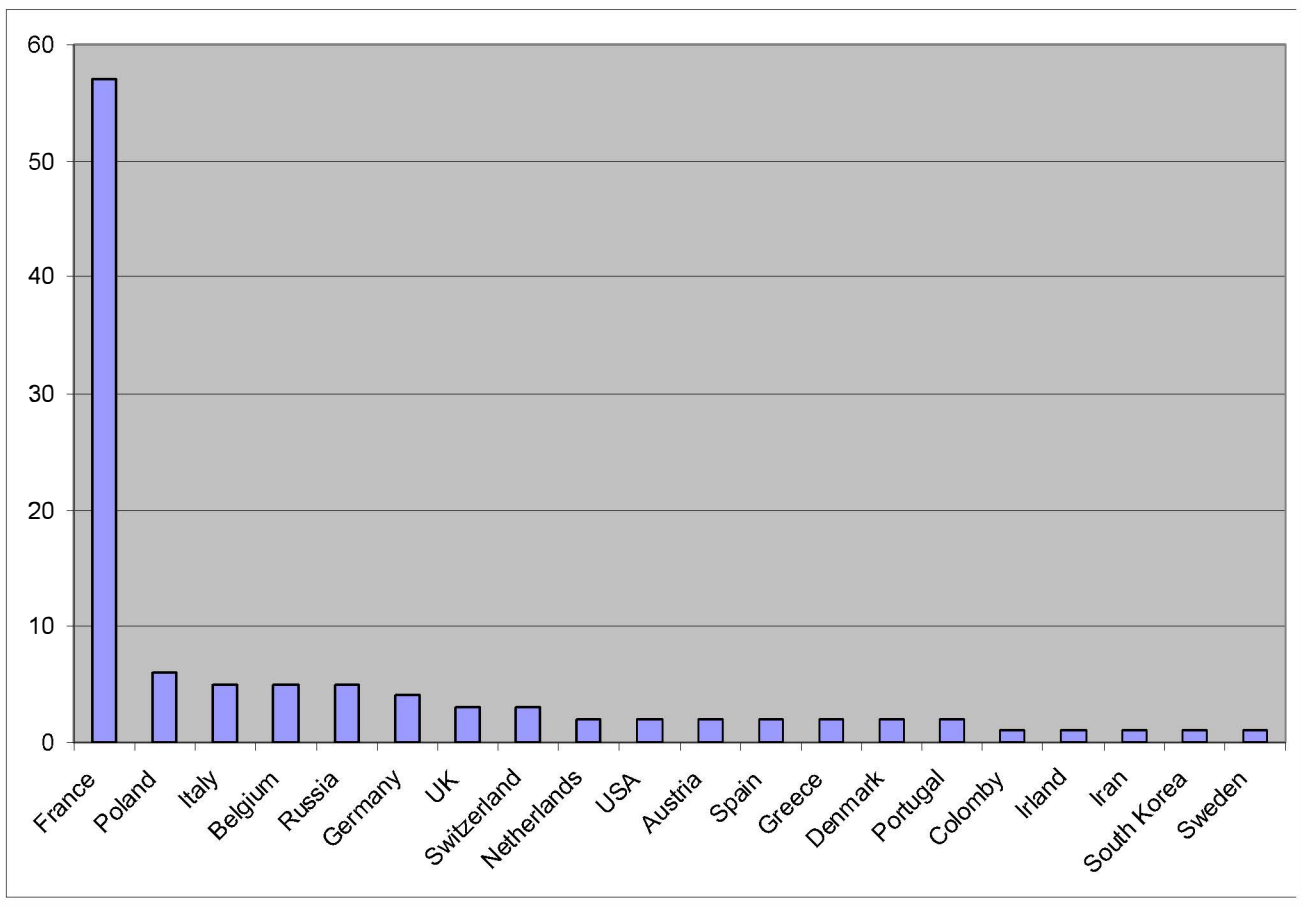

Fig. 5. Country affiliation of the 112 attendees

## IV. Conclusion

There is a rising interest for MRAM in the microelectronics industry. The R&D in this field is accelerating since more and more major microelectronics players are getting active in the field. MRAM and more generally hybrid CMOS/magnetic electronics now clearly appear in microelectronics roadmaps. However, a large cultural gap still subsists between magnetism and microelectronics communities. We should reinforce our efforts to bridge this gap by organizing more joint events gathering experts from the two fields and by introducing magnetism and spinelectronics courses in the initial education of microelectronic engineers. The Introductory Course on MRAM (InMRAM) organized annually in Grenoble, France is an initiative in this direction.

## Acknowledgment

Liesl Folks, Bruce Terris, Ron Goldfarb and Takao Suzuki from the IEEE Magnetic Society are acknowledged for their continuous support in this initiative to foster more relationships between magnetism and microelectronics conferences. This effort is also supported by the European Commission through the ERC Adv grant HYMAGINE (ERC 246942).

## References

[1] M. N. Baibich, J. M. Broto, A. Fert, F. Nguyen Van Dau, and F. Petroff, P. Etienne, G. Creuzet, A. Friederich, and J. Chazelas, « Giant Magnetoresistance of (001)Fe/(001)Cr Magnetic Superlattices », Phys. Rev. Lett. 61, 2472–2475 (1988).

[2] [2] B. Dieny, V. S. Speriosu, S. S. P. Parkin, B. A. Gurney, D. R. Wilhoit, and D. Mauri, “Giant magnetoresistive in soft ferromagnetic multilayers”, Phys. Rev. B 43, 1297–1300 (1991).

[3] [3] J. S. Moodera, Lisa R. Kinder, Terrilyn M. Wong, and R. Meservey, “Large Magnetoresistance at Room Temperature in Ferromagnetic Thin Film Tunnel Junctions”, Phys. Rev. Lett. 74, 3273–3276 (1995).

[4] [4] T.Miyazaki, N.Tezuka, “Spin polarized tunneling in ferromagnet/insulator/ferromagnet junctions “, Journ.Mag.Mag.Mat., 151, 403 (1995).

[5] [5] S.Yuasa, T.Nagahama, A.Fukushima, Y. Suzuki and K. Ando, “Giant room-temperature magnetoresistance in single-crystal Fe/MgO/Fe magnetic tunnel junctions”, Nature Materials 3, 868 - 871 (2004)

[6] [6] S.S.P.Parkin, C.Kaiser, A.Panchula, P.Rice, B.Hughes, M.Samant, S.H.Yang, “Giant tunnelling magnetoresistance at room temperature with MgO (100) tunnel barriers”, Nature Materials, 3 862, (2004).

[7] [7] J.C.Slonczewski, “Current-driven excitation of magnetic multilayers,”, 159, L1, (1996).

[8] [8] L.Berger, “Emission of spin waves by a magnetic multilayer traversed by a current”, Phys.Rev.B, 54, 9353, (1996).

[9] [9] J. A. Katine, F. J. Albert, and R. A. Buhrman, E. B. Myers and D. C. Ralph, “Current-Driven Magnetization Reversal and Spin-Wave Excitations in Co /Cu /Co Pillars”, Phys. Rev. Lett. 84, 3149–3152 (2000).

[10] [10] Y.Huai, F.Albert, P.Nguyen, M.Pakala, and T.Valet, “Observation of spin-transfer switching in deep submicron-sized and low-resistance magnetic tunnel junctions”, Appl. Phys. Lett. 84, 3118 (2004).

[11] [11] http://www.itrs.net/links/2010ITRS/FinalDrafts/FINAL.Report Memory Assessment.pdf

[12] [12] T.Kawahara,” Challenges toward gigabit-scale spin-transfer torque random access memory and beyond for normally off, green information technology infrastructure”, J. Appl. Phys. 109, 07D325 (2011).

[13] [13] I L Prejbeanu, M Kerekes, R C Sousa, H Sibuet, O Redon, B Dieny and J P Nozières, “Thermally Assisted MRAM”, J. Phys.: Condens. Matter 19 165218 (2007).

[14] [14] B.Dieny, R.Sousa, S.Bandiera, M.Castro Souza, S.Auffret, B.Rodmacq, J.P.Nozieres, J.Hérault, E.Gapihan, I.L.Prejbeanu, C.Ducruet, C.Portemont, K.Mackay, B.Cambou, “Extended scalability and functionalities of MRAM based on thermally assisted writing” IEEE Electron Devices Society, Proceedings of the IEDM 2011 conference, Washington DC, Dec2011.